\documentclass[prl,twocolumn,superscriptaddress,amsfonts,amssymb,showpacs,a4]{revtex4}
\usepackage{graphicx} 
\usepackage{dcolumn} 
\usepackage{bm} 
\usepackage[utf8]{inputenc}
\usepackage{amsmath}
\usepackage{amssymb}
\usepackage{times}

\usepackage[usenames]{color}
\usepackage{ulem}

\begin{document}                                                   %

\title{Model potential for the description of metal/organic interface states}

\newcommand{\DIPC}[0]{{
Donostia International Physics Center, 20018 Donostia-San
Sebasti\'an, Spain}}

\newcommand{\CFM}[0]{{
Centro de F\'\i sica de Materiales (CSIC-UPV-EHU) and Materials
Physics Center (MPC), 20018 San Sebasti\'an, Spain}}

\newcommand{\UMR}[0]{{
Fachbereich Physik und Zentrum f{\"u}r Materialwissenschaften,
Philipps-Universit{\"a}t, 35032 Marburg, Germany}}

\author{N.~Armbrust}
\affiliation{\UMR}

\author{F. Schiller}
\affiliation{\UMR}
\affiliation{\CFM}

\author{J.~G{\"u}dde}
\affiliation{\UMR}

\author{U.~H{\"o}fer}
\affiliation{\UMR} \email{Hoefer@physik.uni-marburg.de}

\date{\today}


\begin{abstract}                                       %
 We present an analytical one-dimensional model potential for
the description of electronic interface states that form at the
interface between a metal surface and flat-lying adlayers of
$\pi$-conjugated organic molecules.
 The model utilizes graphene as a universal representation of
these organic adlayers.
 It predicts the energy position of the interface state as well as
the overlap of its wave function with the bulk metal without free
fitting parameters.
 We show that the interface state's energy depends systematically
on the bond distance between the carbon backbone of the adayers
and the metal.
 The general applicability and robustness of the model is
demonstrated by a comparison of the calculated energies with
numerous experimental results for a number of flat-lying organic
molecules on different closed-packed metal surfaces that cover a
large range of bond distances.
\end{abstract}
\maketitle

%
\section{Introduction}                                             %
 The charge transfer at the interface between a metal and a layer of organic molecules plays a decisive role in the functionality of organic semiconductor devices and for future applications
of molecular electronics.
 It depends crucially on the energy alignment and the wave function overlap of electronic states at such interface, which also
governs the binding and even the growth of the
molecular layer~\cite{Koch13,Lindst06cr,Hwang09}.
 These interface states can either originate from localized molecular orbitals of the organic layer or from delocalized electronic states of the metal.
 The latter is characteristic for $\pi$-conjugated organic molecular layers as became first apparent for perylene-tetracarboxylic-acid-dianhydride (PTCDA) on Ag(111) for which an unoccupied, strongly dispersive interface state has been found~\cite{Temirov06nat,Schwalb08prl}.
 A similar interface states has been also observed for the naphthalene-based variant NTCDA on Ag(111)~\cite{Marks11prb1}, for PTCDA on Ag(100)~\cite{Galbra14jpcl} and meanwhile also for a number of other systems~\cite{Tamai08prb, Yamane09jelsp, Scheybal09prb, Ziroff09ss,
Andrews10jcp, Heinrich2011prl, Schmid2011jpcc, Hong2012acsnano,
Faraggi2012jpcc, Wiessner2012njp, Umbach2012jp,  Ilyas2013mp,
Heidorn2013jpcc}.
 Time- and angle-resolved two-photon photoemission (2PPE) experiments on PTCDA/Ag(111) concluded from the dispersion and the rather short inelastic lifetime of this state that it must originate from the Shockley surface state of the bare Ag(111) substrate which is upshifted from below the metallic Fermi level by as much
as 0.7 eV due to the interaction with the molecular layer~\cite{Schwalb08prl,Sachs09jcp}.
 This interpretation was subsequently confirmed by density-functional theory (DFT) calculations \cite{Dyer10njp,Zaitsev10jetp,Marks11prb1,Zaitsev12prb,Tsirkin15prb,Jakob16prb}, which showed that the hybridization of molecular and metallic states is rather small in the region of the projected band gap of the metal.

 It turned out, however, that a realistic description of organic molecules on metal surfaces by DFT is challenging although it is one of the most widely used approaches for the determination of the geometric and electronic structure at surfaces and interfaces~\cite{Pilani13}.
 The large size of the organic molecules does not only require a large supercell within a slab model in the lateral direction.
 A reasonable description of the intrinsic Shockley surface state of the metal makes it necessary to consider also a large number of metal layers~\cite{Zaitsev12prb}.
 Both make such calculations very time-consuming.
 Moreover, metal-organic interfaces require tailored calculations methods~\cite{Jakob16prb}, because conventional DFT neither correctly accounts for van der Waals forces which have an important contribution to the interaction between organic molecules and metal surfaces nor for the correct long-range interaction in front of metal surfaces.

 In order to highlight the main physical mechanism for the formation of the delocalized interface state at organic/metal interfaces without the help of complex DFT calculations, we propose in this letter a one-dimensional description by an analytical model potential that we have recently developed for the description of image-potential states at the center of the Brillouin zone in graphene/metal systems~\cite{Armbrust15njp}.
 We will show that the same model potential does not only predicts the energy of the interface state in graphene/metal systems, but can be also applied to other flat lying molecular layers with a similar $\pi$-$\pi$ interaction as in graphene.
 Our model calculation clearly illustrates how the interface state develops from the former Shockley-type surface state of the bare metal substrate with increasing interaction between the molecular film and the metal.
 By comparing our model results with available experimental data for different organic molecules, we show that our model is able to describe the systematic dependence of the interface state's energy on the bond distance between the carbon backbone and the metal with predictive power if this distance and the work function are known.

\section{Model potential}                                          %
\begin{figure}[t]
    \includegraphics[width = 0.8\columnwidth]{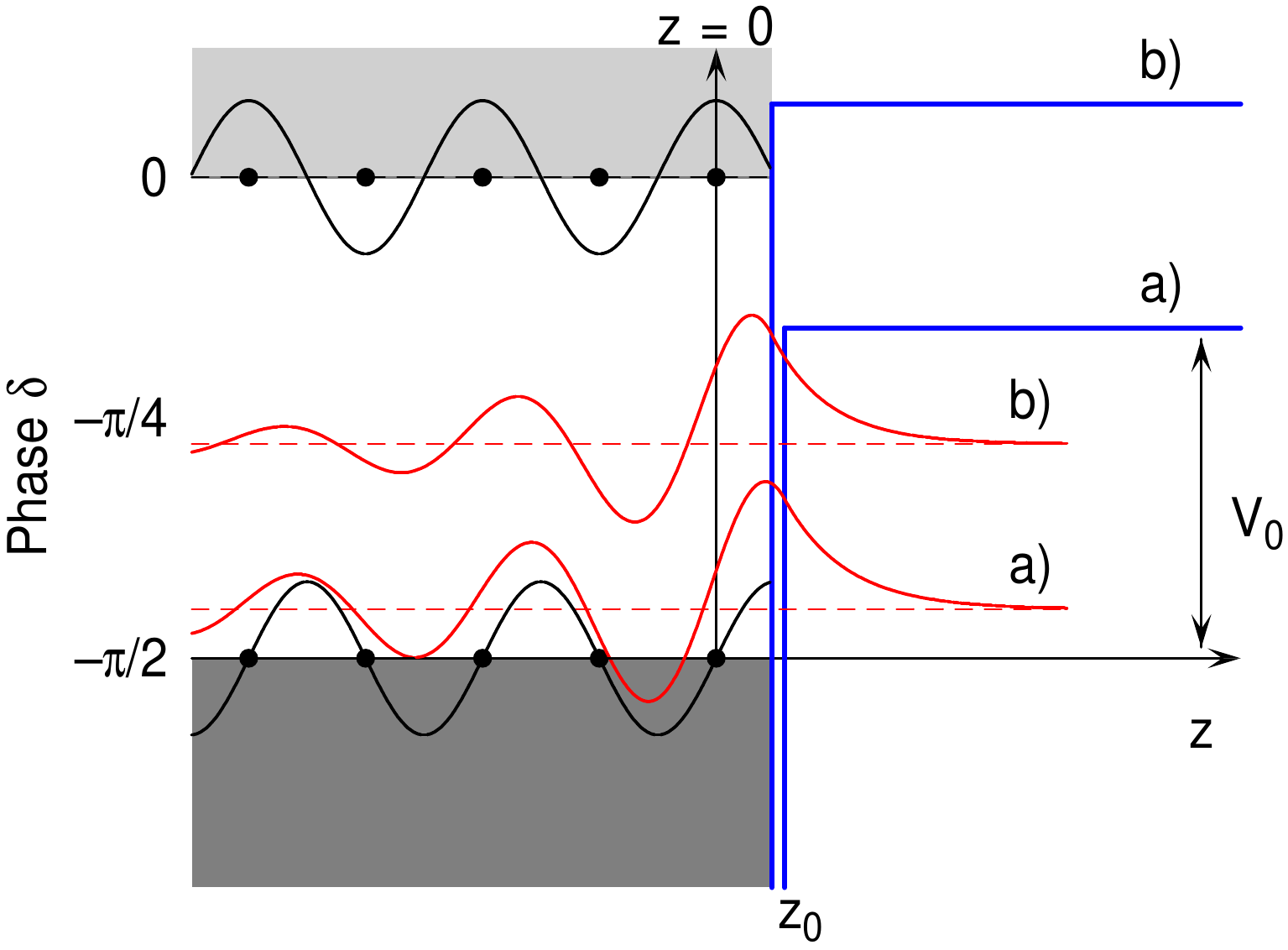}
    \caption[]{Solutions $\Psi(z)$ of the Schr{\"o}dinger equation
        for the Shockley surface state (red solid
        lines) for two different positions $z_{0}$  and heights $V_{0}$
        of the surface barrier (blue solid lines). Black solid lines show the bulk solutions at the top and bottom of the bulk band gap, respectively. Black dots depict the positions of the metal atoms.}
    \label{fig_two_band}
\end{figure}

Our one-dimensional model potential is based on the nearly-free-electron model for the bulk.
 In this approximation, the electronic states within the metal are given by the solution of the Schrödinger equation under the influence of a weak periodic pseudopotential.
 Already within this simple model, the formation of the Shockley surface state can be described by introducing a
potential barrier at the surface.
 For illustration of the basic properties of this surface state, we first recall the textbook example of a step barrier at a distance $z_0$ from the position of the topmost atom at $z=0$~\cite{Maue35zp,Zangwill88}.
 In the direction perpendicular to the surface, the potential is
in this case given by
\begin{equation}
    V(z)= \left\{
    \begin{array}{ll}
    -V_0 + 2 V_g \cos(gz), & z < z_0 \\
                 ~~~~0,   & z > z_0,
    \end{array}
    \right.
    \label{eq_pot_v}
\end{equation}
with the reciprocal lattice vector $g=2\pi/a$ and the distance
between lattice planes $a$.
 With an appropriate choice of the inner potential $V_0$ and the
corrugation $V_g$, this model provides a good approximation of the
energy-momentum dispersion $E(k)$ of electron states for the chosen direction in simple metals such as Al, Ag or Cu which are derived from $sp$ electrons~\cite{Ashcroft76}.
 The model predicts an energy gap $2 V_g$ in which bulk electronic
states are forbidden.
 The bulk eigenfunctions (Bloch states) at the bottom and at the top of this gap correspond to the $k$-value at the Brillouin zone boundary $(k=\frac{g}{2})$, and are simply given by $
\Psi_\pm(z)=e^{i\frac{g}{2}z} \pm e^{-i\frac{g}{2}z}$.
 In case of a repulsive potential $(V_g\!>\!0)$, $\Psi_{+}(z) =
\cos{(\frac{g}{2}z)}$ corresponds to the upper and $\Psi_{-} =
\sin{(\frac{g}{2}z)}$ to the lower band.
 The corresponding eigenvalues are $E_\pm = -V_0 + \frac{\hbar^2}{2m}
(\frac{g}{2})^2 \pm V_g$.

 The potential barrier at the surface leads to an additional solution with an energy within the bulk band gap.
 The corresponding eigenfunction is located at the surface and consists of an exponentially decaying cosine Bloch function in the bulk and a simple decaying exponential on the vacuum side,
\begin{equation}
    \Psi(z)= \left\{
    \begin{array}{ll}
e^{\kappa z}\cos{(\frac{g}{2}z + \delta)}, & z < z_0 \\
e^{-qz}, & z > z_0
    \end{array}
    \right.
    \label{eq_pot_psi}
\end{equation}
with $q=\sqrt{-2 m E}/\hbar$ and $\kappa =
\frac{2m}{\hbar^2}\frac{V_g}{g}\sin(-2\delta)$.
 The phase shift $\delta$ of the bulk cosine function is directly
related to the energy $E$ of the surface state.
 It varies from $\delta = -\pi /2$ at the bottom to $\delta = 0$ at the top of the bulk band gap.
 It is straight forward to show that the wavefunctions in the bulk
and in the vacuum can only be matched for one specific value of the phase $\delta$ or the respective energy~\cite{Zangwill88}.
 The surface state's energy depends on both, the height $V_0$ and the position $z_0$ of the surface barrier.
 It is, however, much more sensitive on the latter.
 Figure~\ref{fig_two_band} illustrates how the energy of the surface state increases for increasing $V_0$ and decreasing $z_0$.
 With parameters for Ag(111) ($a=2.36$~\AA,  $V_g=2.15$~eV, $V_0=9.56$~eV) and $z_0=a/2$, for example, it turns out that changing the surface state's energy by 1~eV requires a change of the barrier height as large as 4~eV, but only a change of $z_0$ by $\sim a/10$.
 The key point for the following discussion is that adlayers of organic molecules modify the distance as well as the height of the barrier, but shift the energy of the surface state predominantly according to their adsorption distance.

 For our model potential we use a more realistic description of the potential barrier at metal surfaces as was introduced by Chulkov \textit{et al.}~\cite{Chulkov99ss2}.
 This approach accounts for the long-range image-potential which is matched to the periodic bulk potential such that the model potential and its derivative is continuous in space.
 By fitting the matching parameters, not only the work function and the energies of the image-potential states, but also the energy of the Shockley surface state on a number of simple and noble metal surfaces can be quantitatively well reproduced~\cite{Chulkov99ss2}.
 Our model potential combines this metal potential with a potential for the molecular adlayer.

 Recently, we have proposed such one-dimensional model potential for a description of the image-potential states of graphene (g) on metal substrates~\cite{Armbrust15njp}.
 We could show that the energy of the image-potential states as well the coupling of their wave functions to the metal bulk systematically varies as a function of the carbon-metal distance $d_{\rm C}$.
 This potential is composed of four parts
\begin{equation}
    V(z) = V_{\rm m}(z) + V_{\rm g}(z-d_{\rm C}) + V_{\rm \Phi}(z) +\delta V(z),
    \label{eq_gpot_v}
\end{equation}
 where $V_{\rm m}(z)$ denotes the metal potential and $V_{\rm g}(z)$ the potential of the $\pi$-conjugated graphene layer.
 The latter is a parameterized analytic expression of the potential proposed by Silkin {\it et al.}~\cite{Silkin09prb}.
 $d_{\rm C}$ is the distance of the carbon atoms in the
graphene layer with respect to the position of the outermost metal atoms located at $z=0$.
 $V_{\rm \Phi}(z)$ and $\delta V(z)$ are corrections that consider
the difference in work function between the bare and the graphene
covered metal and the influence of higher-order image-charges,
respectively.
 Beside the distance and change of the workfunction, $V(z)$ is fixed by the separate properties of the metal and the adlayer and
does not contain further free fitting parameters.
In particular, the metal potential quantitatively reproduce not only the image-potential states, but also the Shockley surface state on the (111) noble metal surfaces.

 We apply this model potential at first to a graphene layer on Ag(111) and then show that the results for this system can be directly related to adlayers of flat-lying organic molecules containing carbon rings that have a similar $\pi$-$\pi$ interaction as in graphene as long as the corresponding carbon-metal distance and work function is considered.
 For the Ag(111) substrate, we use the parameters given in Ref.~\cite{Chulkov99ss2}.
 For the work function of the combined system, we use $\Phi=4.24$~eV at $d_{\rm C}=3.33$~{\AA} as reported in Ref.~\cite{Giovan08prl}.
 $V_{\rm g}(z)$, $V_{\rm \Phi}(z)$ and $\delta V(z)$
are determined as described in Ref.~\cite{Armbrust15njp}.
 Figure ~\ref{fig_gpot_gag111_is}~a) shows the combined potential for an exemplary carbon-metal distance of $d_{\rm C}=5$~\AA.
 The wave functions $\Psi$ and energies $E$ of the Shockley surface state (SS) for bare Ag(111) and the interface state (IS) for graphene covered Ag(111) have been calculated at the center of the Brillouin zone ($\bar{\Gamma}$-point) by solving the one-dimensional Schr{\"o}dinger equation numerically by using Numerov's method.
 We characterize the coupling strength of the interface state to the metal bulk by the the fraction $p$ of its probability density for $z<z_{0} = a/2$.
For bare Ag(111) at the $\overline{\Gamma}$-point, we calculate $p=76.23\%$ and $E_{\rm SS}=-59$~meV relative to the Fermi level.

\section{Results and Discussion}                                   %
\begin{figure}[tb!]
    \includegraphics[width = \columnwidth]{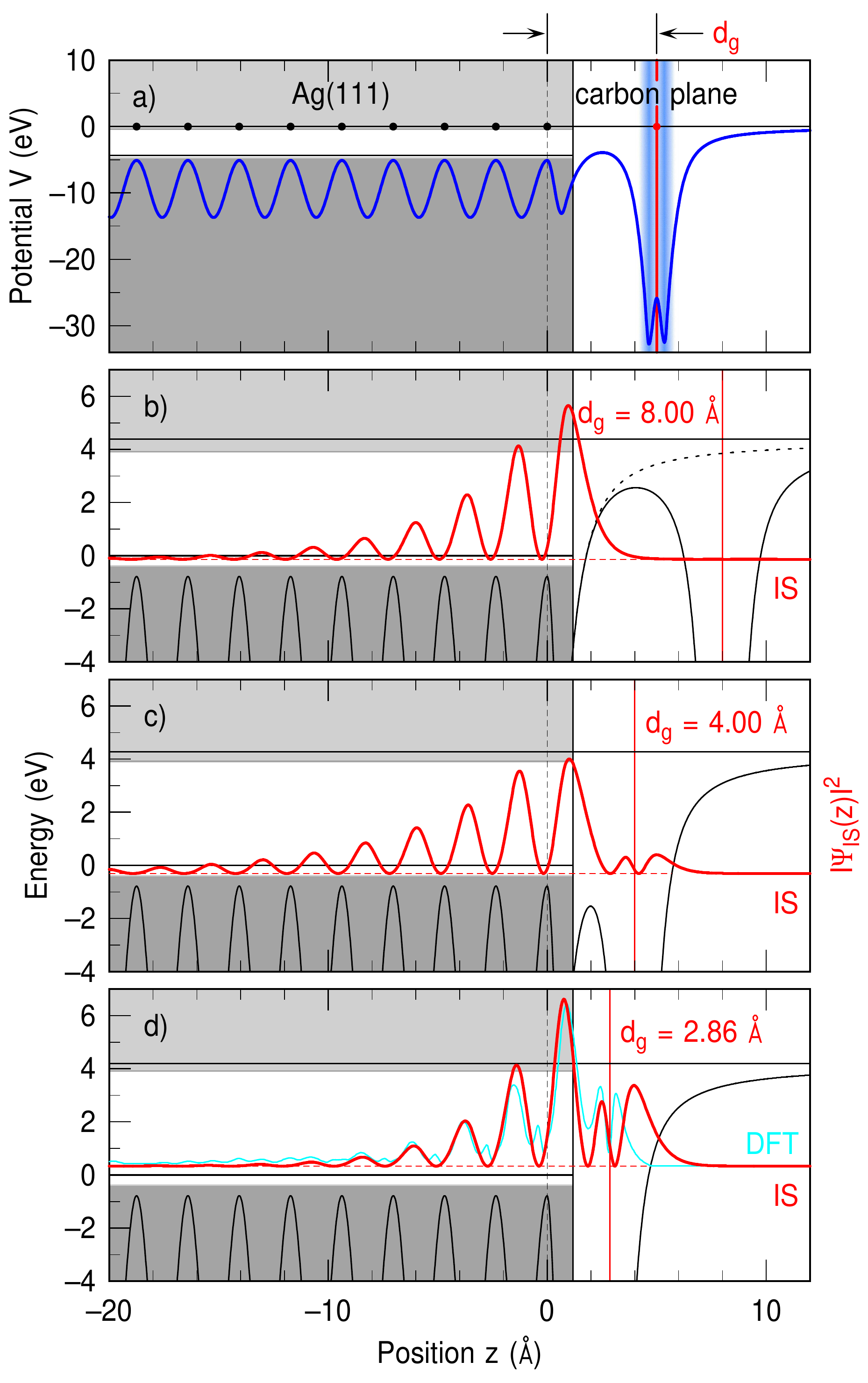}
    \caption[]{(a) One-dimensional model potential $V(z)$ (blue
        solid line) for a carbon layer on Ag(111) at an exemplary
        metal-C distance of $d_{\rm C}=5.0$~\AA.
         The positions of the uppermost Ag atomic layer and the
        carbon layer are depicted by vertical black dashed and red
        solid lines, respectively.
         Black and red circles illustrate the Ag and C atoms, respectively.
        The Ag(111) projected bulk band structure (gray shaded areas)
        has been extended up to the metal surface at $z_{0} = a/2$.
         The blue gradient illustrates the extension of the
        conjugated $\pi$-system of graphene~\cite{Silkin09prb}.
        (b) - (d) show the probability densities
        $|\Psi_{\rm IS}(z)|^{2}$ (solid red curve) of the interface
        state at the graphene/Ag(111) interface for metal-C
        distances $d_{g}$ of $8.00$~\AA~(b), $4.00$~\AA~(c) and
        $2.86$~\AA~(d).
         (b) additionally shows the image-potential of the bare
        metal surface (dotted black line).
         A comparison with the results of the DFT calculations
        for PTCDA/Ag(111) (cyan, data extracted from Fig.~4 of
        \cite{Zaitsev10jetp}) is given in (d)}
    \label{fig_gpot_gag111_is}
\end{figure}

In the following we discuss the transition of the Shockley state of the bare metal into the interface state upon approaching the carbon layer to the Ag(111) surface.
 For this purpose, we have calculated the wavefunction and energy
of the former surface state for different carbon-metal distance $d_{\rm C}$.
 In real systems, $d_{\rm C}$ varies in the range of 2.2--3.7~\AA\ (compare table~\ref{tab_results}) reflecting strong and weak interaction, respectively.
 We start with Fig.~\ref{fig_gpot_gag111_is}~b) at a much larger distance of $d_{\rm C}=8$~{\AA}, where the surface potential of the bare Ag(111) substrate (dotted line) is substantially modified only at distances that are larger than the extension of the Shockley surface state into the vacuum.
 At this distance, the probability density of the interface state (red solid line) is still basically identical with that of the Shockley surface state of bare Ag(111), but its
energy relative to the Fermi level of $E_{\rm IS}=-136$~meV  is slightly reduced.
 This is caused by the reduction of the barrier in the region between the metal and the carbon layer.
 At further decreasing $d_{\rm C}$, there is an interplay between
a further reduction of this barrier and the approach of the barrier between the carbon layer and the vacuum closer to the metal.
 The first leads to a decrease of the interface state's energy,
the latter to an increase.
 At a distance of $d_{\rm C}=4$~{\AA} (Fig.~\ref{fig_gpot_gag111_is}~c), which is still larger than in real systems, one can already observe the transition from
the Shockley surface state of the metal to the actual interface state. Its energy of $E_{\rm IS}=-309$~meV is, however, even further reduced.
 Its probability density leaks further into the vacuum and develops
two small maxima around the position of the carbon layer.
 The overlap with the metal bulk decreases to $p=73.83\%$.
 For a further reduction of $d_{\rm C}$, the approach of the barrier between the carbon layer and the vacuum starts to dominate the effective barrier and results in an upshift of the interface state's energy with decreasing $d_{\rm C}$.
 This case is illustrated in Fig.~\ref{fig_gpot_gag111_is}~d, which shows the calculated probability density for $d_{\rm C}=2.86$~{\AA}.
 This is just the experimental determined distance between the carbon backbone of PTCDA and the metal surface when adsorbed as a flat-lying layer on Ag(111)~\cite{Hausch05prl}.
 At this distance, the calculated energy of the interface state is now shifted substantially above the Fermi level to $E_{\rm IS}=+331$~meV.
 The interface state becomes therefore an unoccupied state.
 The reallocation of the probability density from the metal
surface into the interface and the vacuum region is even stronger
as compared to $d_{\rm C}=4.00$~{\AA}
and the penetration into the metal further decreases to $p=64.70\%$.
 The blue solid line shows for comparison the laterally averaged result of a DFT calculation that has been performed for PTCDA on a nine-layer thick Ag(111) slab~\cite{Zaitsev12prb}.
 The agreement between our results for the one-dimensional model potential and the DFT-result is remarkably good, in particular within in metal and between the metal and the carbon plane.
 Above the carbon plane, however, the DFT result extends much less into the vacuum.
 A part of this difference might be connected to the fact that the DFT calculation does not account for the long-range image-potential that is well described by our model potential.
 On the other hand, our one-dimensional model does not consider
lateral variations of the electronic structure which are most important within the molecular layer.

\begin{figure}[t]
    \includegraphics[width = \columnwidth]{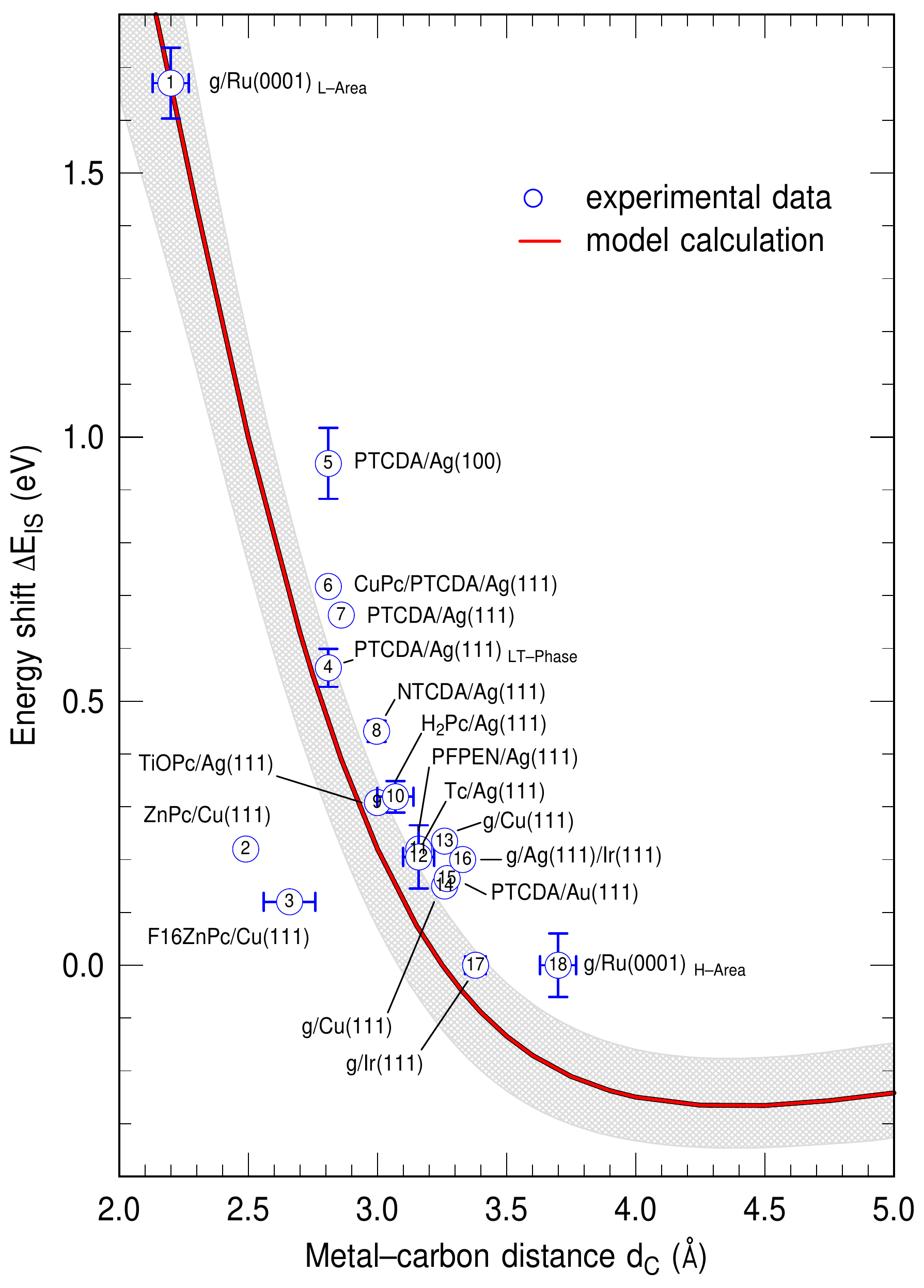}
    \caption[]{Energy shift
    $\Delta E_{\rm IS}$ of the interface state with respect to the energy of the former surface state on the bare metal as a function of the carbon-metal distance $d_{\rm C}$.
     The solid red line shows the calculated results for a carbon layer on Ag(111). The gray area illustrates the variation of these results when changing the work function by $\pm$1~eV. Symbols denote the experimental data listed in table~\ref{tab_results}.}
    \label{fig_is_shift_metals_exp_gpot8}
\end{figure}

 For a comparison with experimental results, we list in Tab.~\ref{tab_results} available data on the energy difference $\Delta E_{\rm IS}=E_{\rm IS}-E_{\rm SS}$ between the interface state and the surface state on the carbon-metal distance $d_{\rm C}$ for a variety of adlayers of organic molecules on metal surfaces.
 In order to emphasize the prototypic character of a graphene layer on a metal surface for these systems, we added also data of g/metal systems.
 If available, $d_{\rm C}$ has been taken from x-ray standing wave
experiments, otherwise from DFT, low energy electron diffraction
or surface x-ray diffraction.
 Fig.~\ref{fig_is_shift_metals_exp_gpot8} depicts the correlation between $\Delta E_{\rm IS}$ and $d_{\rm C}$ for the experimental and the calculated results.
 The red solid line shows the calculated results for a work function of the combined g/Ag(111) system of $\Phi=4.24$~eV at $d_{\rm C}=3.33$~{\AA}~\cite{Giovan08prl}.
 Since the work function difference between the covered and bare metal surfaces varies substantially for the different systems even at comparable $d_{\rm C}$, we depict with the gray areas the variation of the calculated results when changing the work function by $\pm 1$~eV.
 Obviously, the work function, i.e. the height of the potential barrier, has only a minor influence on the energy of
the interface state.
 It depends instead much more sensitive on $d_{\rm C}$ and shows a strong increase for distances below $\sim 3.25$~{\AA}.
 Our model calculation can well reproduce this trend for the majority of the experimental data although the model slightly underestimates the absolute value of the energy shift for most data points.
 We can not confirm the predicted downshift of the interface state at large $d_{\rm C}$ because this regime is not covered by real systems.
 The agreement between model calculation and experiment is particularly good for system with rather weak interaction as, for example, PTCDA and NTCDA on Ag(111).
 In these two cases, our model agrees also very well with the results of a DFT calculation~\cite{Zaitsev12prb}.

 Systems with stronger bonds of the functional groups of the organic molecules typically show a pronounced bending of the molecules towards the surface which reduces the carbon-metal distance at the edges~\cite{Bauer12prb}.
 Because we relate $d_{\rm C}$ to the center of the carbon backbone, this might explain the larger deviation between model and experiment for PTCDA/Ag(100).
 For this system, also DFT calculations~\cite{Galbra14jpcl} clearly
underestimate  $\Delta E_{\rm IS}$ (c.f. Tab.~\ref{tab_results}).
 Even if our simple one-dimensional model gives a reasonable explanation of the main physical mechanism for the interface state formation in a number of systems, it can, however, not account for more complex chemical interactions between the substrate and the molecules.
 This might explain the more pronounced deviations between model and experiment for the phthalocynaines ZnPc/Cu(111) and F16ZnPc/Cu(111) that are subject to a stronger corrugation or distortion of the molecular layer on the substrate~\cite{Yamane10prl}.

\begin{table}[h]
    \caption[]{Experimental data on the carbon-metal distance $d_{\rm C}$ and the energy shift $\Delta E_{\rm IS}$ between the interface state and the surface state for single layers of flat-lying organic molecules and graphene layers on metal surfaces.
 The numbering corresponds to that of the data points in Fig.~\ref{fig_is_shift_metals_exp_gpot8}
 In addition, results of DFT calculations are listed for NTCDA and PTCDA on Ag surfaces.}
    \label{tab_results}
    \begin{tabular*}{\columnwidth}{@{\extracolsep\fill}r r D{.}{.}{9} D{.}{.}{8}}
        \hline\hline
        \multicolumn{1}{c}{\#} &%
        \multicolumn{1}{c}{System} &%
        \multicolumn{1}{c}{$d_{\rm C}$~({\AA})} &%
        \multicolumn{1}{c}{$\Delta E_{\rm IS}$~(eV)}\\\hline\\[-0.7em]

         2 & ZnPc/Cu(111)              & 2.49(3)\mbox{~\cite{Yamane10prl}}    &  0.22\mbox{~\cite{Yamane10prl}}                            \\
         3 & F16ZnPc/Cu(111)           & 2.66(10)\mbox{~\cite{Yamane10prl}}   & \approx\!0.12\mbox{~\cite{Yamane10prl}}                    \\
         4 & PTCDA/Ag(111) (LT-Phase)  & 2.81(2)\mbox{~\cite{Hausch10prb}}    &  0.56(3)\mbox{~\cite{Marks11prb1}\footnotemark[2]}         \\
         5 & PTCDA/Ag(100)             & 2.81(2)\mbox{~\cite{Bauer12prb}}     &  0.95(7)\mbox{~\cite{Galbra14jpcl}\footnotemark[1]}        \\
         6 & CuPc/PTCDA/Ag(111)        & 2.81\mbox{~\cite{Stadtm12prl}}       &  0.72\mbox{~\cite{ZimmerCuPcAg111tbp}\footnotemark[2]} \\
         7 & PTCDA/Ag(111)             & 2.86(1)\mbox{~\cite{Hausch05prl}}    &  0.66\mbox{~\cite{Schwalb08prl}\footnotemark[2]}           \\
         8 & NTCDA/Ag(111)             & 3.00(2)\mbox{~\cite{Stadler07njp}}   &  0.44(2)\mbox{~\cite{Marks11prb1,Jakob16prb}\footnotemark[2]}         \\
         9 & TiOPc/Ag(111)             & 3.00(3)\mbox{~\cite{Kroger10dr}}     &  0.31\mbox{~\cite{SchillerTiOPcAg111tbp}\footnotemark[2]} \\
        10 & H2Pc/Ag(111)              & 3.07(7)\mbox{~\cite{Kroger12prb}}    &  0.32(3)\mbox{~\cite{Caplins14jpcl}}                       \\
        11 & Tc/Ag(111)                & 3.16\mbox{~\cite{Zaitsev16arxiv}} &  0.22\mbox{~\cite{Soubatch11prb}}                          \\
        12 & PFPEN/Ag(111)             & 3.16(6)\mbox{~\cite{Duhm10prb}}      &  0.21(6)\mbox{~\cite{SchillerTiOPcAg111tbp}\footnotemark[2]} \\
        15 & PTCDA/Au(111)             & 3.27(2)\mbox{~\cite{Henze07ss}}      &  0.164(4)\mbox{~\cite{Ziroff09ss}}                         \\\\[-0.7em]\hline\\[-0.7em]
         1 & g/Ru(0001) (L-Area)       & 2.20(7)\mbox{~\cite{Armbrust12prl, Wang08prlComm, Wang08pccp, Moritz10prl}}  &  1.67(7)\mbox{\cite{Armbrust12prl}\footnotemark[1]} \\
        13 & g/Cu(111)                 & 3.26\mbox{~\cite{Giovan08prl}}       &  0.24\mbox{~\cite{Pagliara15prb}\footnotemark[2]}          \\
        14 & g/Cu(111)                 & 3.26\mbox{~\cite{Giovan08prl}}       &  0.15\mbox{~\cite{Jeon13ns}}                             \\
        16 & g/15ML-Ag(111)/Ir(111)    & 3.33\mbox{~\cite{Giovan08prl}}       &  0.20\mbox{~\cite{Jolie15prb}}                             \\
        17 & g/Ir(111)                 & 3.38(4)\mbox{~\cite{Jean15prb}\footnotemark[3]}      &  0.00\mbox{~\cite{Niesner12prb}}           \\
        18 & g/Ru(0001) (H-Area)       & 3.70(7)\mbox{~\cite{Armbrust12prl, Wang08prlComm, Wang08pccp, Moritz10prl}} &  0.00(6) \mbox{\cite{Armbrust12prl}\footnotemark[1]} \\\\[-0.7em]\hline\\[-0.7em]
           & PTCDA/Ag(100) (DFT)       & 2.81(2)\mbox{~\cite{Bauer12prb}}     &  0.63\mbox{~\cite{Galbra14jpcl}\footnotemark[1]}           \\
           & PTCDA/Ag(111) (DFT)       & 2.86(1)\mbox{~\cite{Henze07ss}}      &  0.56\mbox{~\cite{Zaitsev12prb}}                           \\
           & NTCDA/Ag(111) (DFT)       & 3.00(2)\mbox{~\cite{Stadler07njp}}   &  0.32\mbox{~\cite{Zaitsev12prb}}                           \\
        \hline\hline
    \end{tabular*}
    \footnotetext[1]{relative to energy of the respective surface
        resonance of the bare substrate}
    \footnotetext[2]{relative to energy of Shockley state
        of the bare substrate from Ref.~\cite{Reinert01prb}}
    \footnotetext[3]{average distance}
\end{table}

\section{Conclusion}                                               %
 We have shown that a parameterized one-dimensional analytical
model potential is able to explain the main physical mechanism for the formation of the electronic state that arises at the interface between a monolayer of graphene-like organic molecules and a metal substrate.
 In particular, we could show that this interface state originates from the former Shockley-type surface state of the bare metal substrate that is subject to a strong upshift in energy upon adsorption of the molecular adlayer, which crucially depends on the carbon-metal distance.
 The model makes it possible to reproduce and predict the binding energy of this state for a number of organic/metal interfaces
with reasonable accuracy if the carbon-metal distance and the corresponding work function are known.

\section*{Acknowledgement}                                         %
 We gratefully acknowledge funding by the Deutsche
Forschungsgemeinschaft through SFB 1083 {\sl Structure and Dynamics
of Internal Interfaces}, Project B6.


\section*{References}                                              %
\bibliographystyle{prsty}

\end{document}